\documentclass{pasj00}
\draft

\begin{document}
\SetRunningHead{T.Enoto et al.}{Running Head}

\title{
Soft and Hard X-Ray Emissions from \\
 the Anomalous  X-ray Pulsar 4U 0142+61 \\
 Observed with Suzaku
}



%
\author{%
T. Enoto\altaffilmark{1,2}, 
K. Makishima\altaffilmark{1,3}, 
K. Nakazawa\altaffilmark{1}, \\
M. Kokubun\altaffilmark{4},
M. Kawaharada\altaffilmark{4}
J. Kotoku,\altaffilmark{5},
and
N. Shibazaki\altaffilmark{6}
}

\email{enoto@stanford.edu}

\altaffiltext{1}{
Department of Physics, The University of Tokyo,
7-3-1 Hongo, \\
Bunkyo-ku, Tokyo 113-0033, Japan
}

\altaffiltext{2}{
Current address: Kavli Institute for Particle Astrophysics and Cosmology, \\
Department of Physics and SLAC National Accelerator Laboratory,  \\
Stanford University, Stanford, CA 94305, USA
}

\altaffiltext{3}{
High Energy Astrophysics Laboratory, 
Institute of Physical and Chemical Research (RIKEN),  \\
Wako, Saitama 351-0198, Japan
}

\altaffiltext{4}{
Institute of Space and Astronautical Science (ISAS), 
Japan Aerospace Exploration Agency (JAXA),\\
3-1-1 Yoshinodai, Chuo-ku, Sagamihara, Kanagawa 252-5210, Japan
}


\altaffiltext{5}{
Department of Radiological Technology, Faculty of Medical Technology, Teikyo University \\
2-11-1 Kaga, Itabashi-ku,  Tokyo 173-8605, Japan
}

\altaffiltext{6}{
Department of Physics, Faculty of Science, Rikkyo University, \\
3-34-1 Nishi-Ikebukuro, Toshima-ku, Tokyo 171-8501, Japan
}


\maketitle

\begin{abstract}
The anomalous X-ray pulsar 4U 0142+61 was observed with Suzaku
	on 2007 August 15 for a net exposure of $\sim$100 ks,
	and was detected in a 0.4 to $\sim 70$ keV energy band.
The intrinsic pulse period was determined as 
	$ 8.68878 \pm 0.00005 $ s,
	in agreement with an extrapolation from previous measurements.
The broadband Suzaku spectra enabled 
	a first simultaneous and accurate measurement 
	of the soft and hard components of this object by a single satellite.
The former can be reproduced by two blackbodies,
	or slightly better by 
	a resonant cyclotron scattering model.
The hard component can be approximated by a 
	power-law of photon index $\Gamma_{\rm h} \sim 0.9$
	when the soft component is represented by the resonant cyclotron scattering model,
	and its high-energy cutoff is constrained as $>180$ keV.
Assuming an isotropic emission at a distance of 3.6 kpc,
	the unabsorbed 
1--10 keV and 10--70 keV luminosities
	of the soft and hard components are  calculated as 
	$2.8\times 10^{35}$ erg s$^{-1}$ and $6.8\times 10^{34}$ erg s$^{-1}$, 
	respectively.
Their sum becomes 
	$\sim 10^3$ times as large as 
	the estimated spin-down luminosity.
On a time scale of  30 ks,
	the hard component exhibited evidence of  variations
	either in its normalization or pulse shape.
\end{abstract}

\section{Introduction}
Anomalous X-ray pulsars (AXPs),
	comprising at present $\sim$9 objects discovered in the local universe,
	are characterized by
	rotational periods in the range $P=5-12$ s, 
	and 
	spin down rates as $\dot{P} \sim 10^{-11}$ s s$^{-1}$.
Together with soft gamma repeaters (SGRs),
	AXPs are thought to form 
	a subgroup of neutron stars called  ``magnetars'',
	of which the surface magnetic field strengths 
	are believed to reach $\sim 10^{14-15}$ G
	\citep{Thompson95,Thompson96,Woods06}.
In energies below $\sim$ 10 keV,
	AXPs exhibit spectra which are much softer 
	than those of ordinary binary X-ray pulsars,
	while harder than those of isolated neutron stars.
Their X-ray luminosities ($10^{34-35}$ erg s$^{-1}$),
	which often exceed up to two order of magnitude 
	those available from their spin-down power ($\sim$$10^{32}$ erg s$^{-1}$),
	are usually thought to be sustained by a release
	of energies stored in  their ultra-strong magnetic fields 
	\citep{Duncan92,Thompson95}.

A novel observational window onto magnetars
	has been opened by the INTEGRAL 
	discoveries of a distinct pulsed hard X-ray component
	from at least 3 AXPs and 2 SGRs \citep{Kuiper04,2005A&A...433L...9M,denHartog2006,Kuiper06,2007Ap&SS.308...51G}.
This  component extends to $\sim 100$ keV or more
	with a very flat photon index of $\Gamma \sim 1$, 
	and  
	exhibits a luminosity comparable to or higher 
	than that in the softer X-ray band.
Although theoretical accounts for this enigmatic component are 
	far from settled (e.g., \cite{Heyl05,Beloborodov07,Baring07}),
	its near-absence in other types of X-ray sources suggests 
	its close relation to the proposed strong magnetic fields
	of magnetars.

The study of magnetars now requires a broad energy band,
	because their soft and hard components,
	which are generally variable, 
	must be measured simultaneously and accurately.
This makes Suzaku \citep{Mit07} an ideal observatory.
In fact, 
	the X-ray Imaging Spectrometer (XIS),
	operating in the 0.2--12 keV range \citep{Koyama07},
	provides high-quality data of their soft components.
Simultaneously, 
	the Hard X-ray Detector (HXD;  \cite{Takahashi07,Kokubun07}),
	consisting of HXD-PIN (10--70 keV) and HXD-GSO (40--600 keV),
	can conduct detailed spectroscopic studies of their hard components
	in considerably shorter exposures than are needed by INTEGRAL.

Such wide-band spectroscopic observations with Suzaku 
	have allowed 
	detections of the hard-tail component from activated magnetars,
	including SGR~0501+4516 \citep{Enoto09ApJL, Rea2009MNRAS, Enoto2010ApJ}
	and 1E 1547.0-5408 \citep{Enoto2010PASJ}.
Through these observations,		
	the hard X-ray emission from SGR~0501+4516
	was revealed to form a component distinct from the soft X-rays,
	while the power-law-shaped hard-tail component of 1E~1547.0-5408
	to extend to the soft X-ray band below 10 keV.
Thus, a key point of the observation of magnetars is
	to detect and quantify their wide-band spectra simultaneously, 
	without hampered by cross-calibration uncertainties or non-simultaneity 
	 involved in multi-satellite observations.
This is a great advantage provided with Suzaku.

\begin{figure}[hbt]
\begin{center}
\FigureFile(130mm,80mm){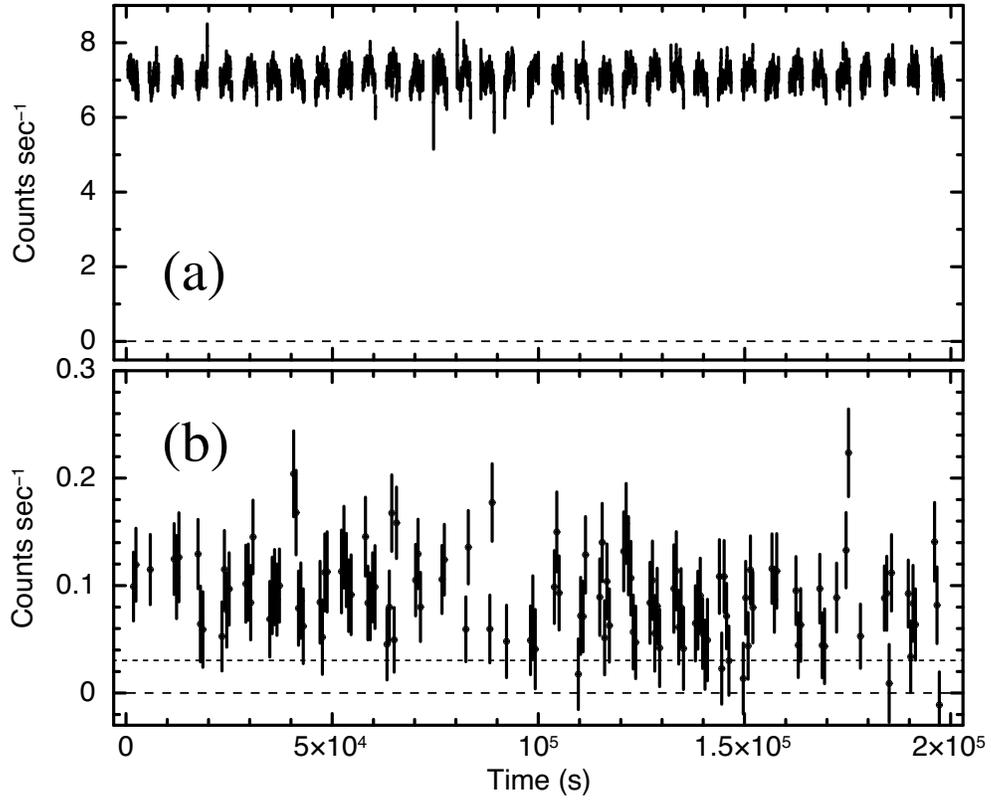}
\end{center}
\caption{
(a)
Background-subtracted 0.4--10 keV light curve of XIS3, 
	with a binning of 160 s.
(b) 
The NXB-subtracted HXD-PIN light curve 
	in the 10--70 keV range, with 550 s binning.
It still includes the CXB contribution by  $\sim$0.03 c s$^{-1}$,
	which is shown as a dotted line.
}
\label{fig:lc}
\end{figure}

\begin{figure}[htb]
  \begin{center}
    \FigureFile(140mm,140mm){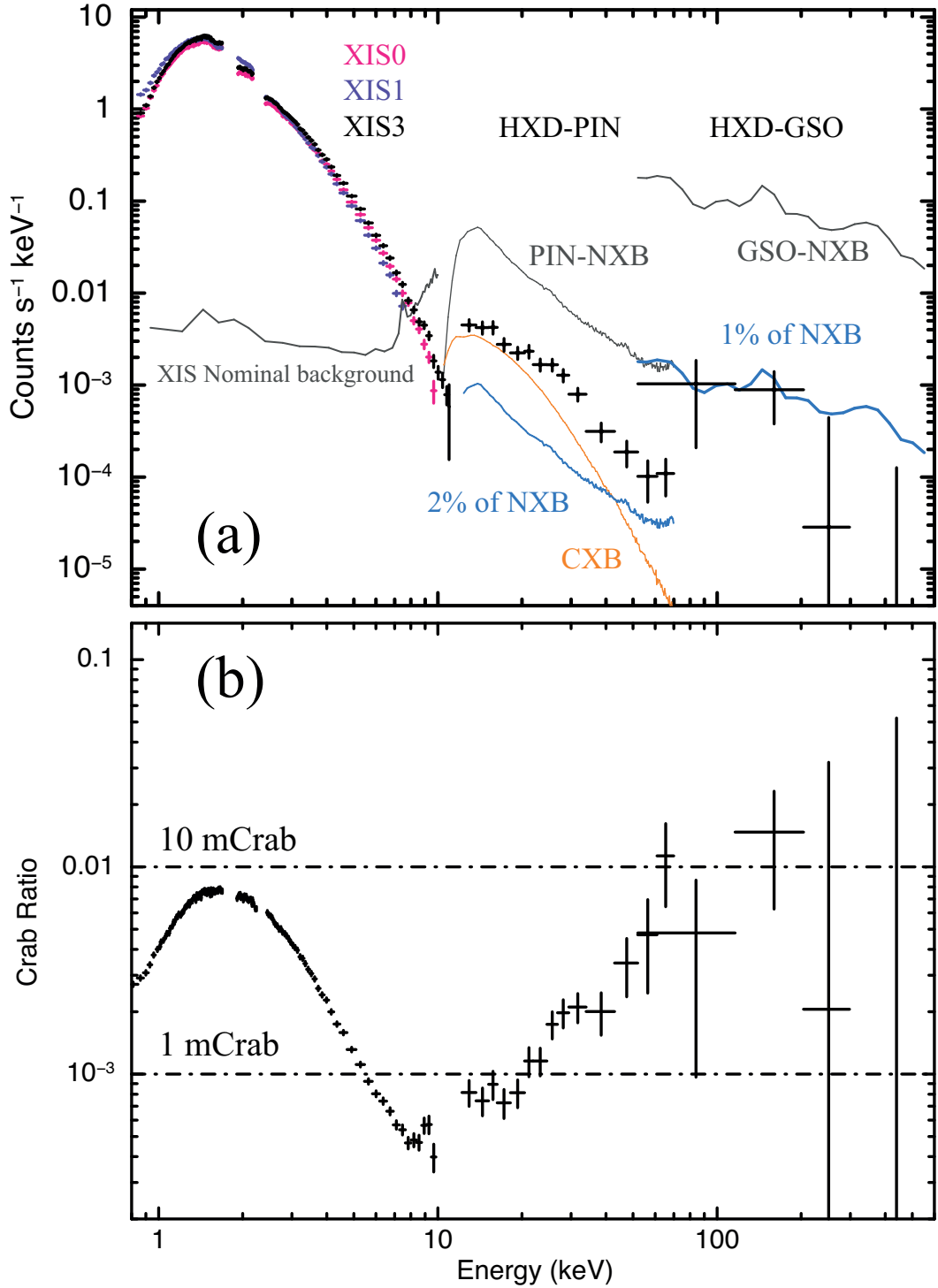}
  \end{center}
  \caption{
(a) 
Background-subtracted Suzaku spectra of 4U 0142+61 
	obtained with the XIS, HXD-PIN, and HXD-GSO.
Not only the NXB but also 
	the CXB contribution was subtracted from the HXD-PIN data.
The error bars are statistical only for the XIS and HXD-PIN,	
	while those of HXD-GSO include 
	systematic errors as well.
\ (b) 
The same spectra as presented in panel (a),
	but divided by those of the Crab Nebula.
}
\label{fig:rawspec}
\end{figure}

In the present paper,
	we report on a high sensitivity  broadband observation 
	of a prototypical magnetar 4U 0142+61 made with Suzaku.
This object is one of the most luminous AXPs 
	with $P$$\sim$8.68 s \citep{Israel94},
	located at the Galactic anti-center direction,
	probably on the Perseus arm \citep{Durant06a}.
{
Its pulsed hard X-ray emission was detected 
	up to $\sim 200$ keV by INTEGRAL and RXTE
	\citep{Kuiper06,denHartog2006,Hartog08}.
On the other hand, 
	past COMPTEL observations gave a flux upper limit 
	in energies above 750 keV \citep{Kuiper06, denHartog2006}.
If the source exhibits little time variations between these observations,
	the hard X-ray spectrum must turn over at energies between 
	$\sim$200 and $\sim$750 keV, 
	though it has not yet been clearly detected.

\section{Observation and Data Reduction}

We observed 4U 0142+61 with Suzaku
	from 04:04 UT on 2007 August 13 to 12:30 UT on August 15,
	for a gross duration of  203 ks
	and a net exposure of $\sim 100$ ks.
The XIS and the HXD were both operated in their normal modes, 
	except that 1/4 window option was applied to the XIS
	to achieve a time resolution of 2 s.

We placed the source at the HXD nominal position,
	and adjusted the roll angle 
	to avoid contaminating bright hard X-ray sources. 
In particular, 
	we tried to minimize the confusion with 
	the Be X-ray binary RX J0146.9+6121.
Located $24'$ off 4U 0142+61,
	this source exhibited 
	a 20--50 keV flux comparable to that of 4U 0142+61 
	during INTEGRAL observations in 2003-2004 
	\citep{denHartog2006}.
As a result of the roll-angle adjustment, 
	the HXD-PIN effective area onto 
	RX J0146.9+6121 was $\lesssim$29 \% of that 
	onto 4U 0142+61 in the 10-70 keV energy band.
Based on this, and the reported steep spectrum 
	($\Gamma \sim$3; \cite{denHartog2006}),
	we regard 
	the contamination of RX J0146.9+6121 to the HXD negligible.
There were no other catalogued bright 
	X-ray sources inside the HXD field of view.

We used the Suzaku data prepared via version 2.1 processing,
	and the HEASOFT version 6.4 or later tools.
The employed data screening criteria were;
	(a) 
	the time after an exit from the South Atlantic Anomaly  should be
	$>$500 s and $>$436 s for the HXD and the XIS, respectively,
	while the time to the next entry should be $>$180 s for the HXD.
	(b) the target should be above the Earth rim by $>$5$\degree$ for the HXD and the XIS,
	and,
	in case of the sunlit Earth rim, that should be $>$20$\degree$ for the XIS;
	(c) the instantaneous pointing direction should be within $1'.5$ of the mean;
	and
	(d) the geomagnetic cutoff rigidity should be $>$6 GV.
As a result of these screenings, 
	we archived a net exposure of 
	99.7 ks with the XIS, and 94.7 ks with the HXD.

\section{Data Analysis and Results}

\subsection{Background subtraction}
\label{sub:bgsubt}
We accumulated the screened data of the XIS 
	from a region within $2'.0$ (2.8 mm) radius of the source centroid, 
	and derived a background spectrum from another region 
	which is symmetric with respect to the XIS field-of-view center.
Corresponding rmf and arf files were created using 
	{\tt xisrmfgen} and {\tt xissimarfgen} tools
	\citep{Ishisaki2007PASJ}, respectively.

The non-Xray background (NXB) of the HXD must be subtracted 
	using appropriate models that emulate NXB events 
	to be observed during the on-source exposure.
Specifically, we utilized 
	LCFITDT (bgd\_d 2.0 ver0804) and LCFIT (bgd\_d 2.0 ver0804) methods 
	\citep{Fukazawa2009PASJ} 
	to create the fake NXB events of HXD-PIN and HXD-GSO, respectively.
These fake photons were also screened 
	by the same criteria as for the on-source event data.

Using the data acquired during Earth-occultation periods 
	($\sim$20 ks in total) in the present observation,
	we confirmed that the HXD-PIN and HXD-GSO spectra obtained therein 
	can be reproduced by the employed NXB models
	both within $\sim$1\%,
	which agrees with the typical uncertainties of these models.
This calibration suggests that we are slightly ($\sim$0.6\%)
	under-subtracting the NXB of HXD-GSO.
Thus, we finally subtracted the HXD-GSO background after increasing it by 0.6\%,
	and include systematic uncertainties of 0.6\% 
	after \citet{Fukazawa2009PASJ}.


\subsection{Source detection}
\label{sub:detection}
Figure~\ref{fig:lc} shows 
	background-subtracted light curves of XIS3 and HXD-PIN.
For the soft X-ray light curve, we utilized the XIS3 data
	because those from XIS0 or XIS1  are more 
	subject to artificial intensity variations
	synchronized with the orbital period of Suzaku,
	caused by thermal distortions of  the spacecraft structure \citep{2008PASJ...60S..35U}.
We do not show a light curve of HXD-GSO,
	since the NXB model for HXD-GSO is valid only
	for integrations longer than $\sim$100 ks
	for such low count rate objects.
Thus, 
	the source has been detected clearly 
	not only by XIS3 but also by HXD-PIN,
	as the average HXD-PIN count rate,
	$\sim 0.08$ c s$^{-1}$, significantly exceeds the expected CXB rate 
	of $\sim 0.03$ c s$^{-1}$ \citep{Kokubun07}.

At the binning as shown in Figure~\ref{fig:lc},
	the XIS3 count rate was constant 
	during the observation within 15\%.
The possible variations seen in figure~\ref{fig:lc}a
	can be mostly attributed to slight losses of signal photons, 
	due to the thermal distortion of the spacecraft structure
	which caused  the image outskirts to partially  fall outside the 1/4 window.
The HXD-PIN signals,
	with 550 s binning, 
 are also roughly constant: 
	the NXB-subtracted light curve can be 
	fitted by a constant with $\chi^2/\nu=154.9/126$.
However, 
	the HXD-PIN light curve may be slightly decreasing with time,
	because it can be fitted better ($\chi^2/\nu=142.2/125$) by a linear function
	where the count rate at the end of the observation
	is 35\% lower than that at the beginning.
Both the  XIS3 (2-s bin) and HXD-PIN (1-s bin) counts 
	are consistent with obeying Poisson distributions,
	without any significant short-term increases of their count rates.

Figure~\ref{fig:rawspec}a shows 
	background-subtracted spectra of 4U 0142+61 
	obtained with three XIS cameras, HXD-PIN, and HXD-GSO.
The background levels are also shown.
From the HXD-PIN spectrum, 
	we subtracted not only the modeled NXB described in section~2,
	but also the  CXB spectrum  taken from \citet{Moretti2009A&A}. 
Although the CXB brightness 
	within the HXD-PIN field of view
	is expected to vary by $\sim \pm$12\% from sky to sky \citep{Fukazawa2009PASJ},
	the effect is
	of the order of 0.5\% of the NXB,
	and hence is negligible compared to the NXB modeling uncertainty.
The CXB contribution itself is negligible in the HXD-GSO data.

As seen in figure~\ref{fig:rawspec}a,
	the HXD-PIN signals (after removing the NXB and CXB)
	are statistically significant over the entire 10--70 keV band,
	and stay $\gtrsim 10\%$ of the NXB level therein.
Since the  NXB can be reproduced typically down to $\lesssim$2\%  \citep{Fukazawa2009PASJ},
	the HXD-PIN detection is highly significant with $\gtrsim$3$\sigma$
	even considering the systematic errors.
In fact, 
	this NXB modeling uncertainty is 
	considerably smaller than the statistical errors.	
In contrast, 
	the NXB-subtracted HXD-GSO signals  are found
	at a level of $\sim 1\%$ of the NXB which is  a typical systematic uncertainty.
Therefore, the source detection with HXD-GSO is considered insignificant	
	only from the spectrum.

In order to grasp basic properties of the spectra of 4U 0142+61 
	in a model independent way,
	we normalized them to those of the Crab Nebula 
	(a power-law with a photon index $\Gamma=2.1$).
We utilize the HXD data of the Crab Nebula obtained on 2007 March 20,
	while employ the simulated Crab Nebula spectrum for the XIS,
	in order to avoid pile-up effects at the source centroid.	
The derived ``Crab ratio'' is presented in figure 1b.
Thus, in addition to the long-known soft component, 
	the hard component (section 1) is very clearly visible, 
	rising up from $\sim$10 keV and reaching 10 mCrab at 100 keV.

\subsection{Source pulsations}
\label{sub:pulsations}

After applying barycentric corrections
	to the arrival times of individual photons  \citep{Terada2008PASJ},
	we searched the background-inclusive XIS and HXD data 
	for the 8.7-s source pulsation via standard periodogram analyses.
Since the spacecraft wobbling is insignificant 
	on the relevant  time scales,
	the data from XIS0 and XIS1 ware also incorporated.

Figure~\ref{fig:periodogram} show periodograms 
	obtained with the XIS and HXD-PIN 
	in the 0.8--10 keV and 12--40 keV energy range, respectively.
Thus, the pulsation has been detected clearly by the XIS,
	at a barycentric period of
\begin{equation}
P=8.68878\pm 0.00005 {\rm \ s.}
\label{eq:period}
\end{equation}
This agrees, within 11 $\mu$s, 
	with a value extrapolated from previously
	measured $P$ and 
	the period change of $\dot{P}=1.960\times 10^{-12}$ s s$^{-1}$
	\citep{Gavriil2002}.
The width of the XIS periodgram peak is
	$\sim 3\times 10^{-4}$ s, or $\sim 4\times 10^{-5} P$.
This is consistent with the gross time span of our observation,
	$\sim 2.0\times 10^5$ s (section 2),
	wherein $\sim 3\times 10^4$ pulses are contained.
	
In the 12--40 keV  HXD-PIN periodgram, 
	a peak with $\chi^2/\nu=35.9/15=2.4$ 
	is found at a consistent period.
This gives a chance probability of $8.4\times 10^{-4}$ 
	for the HXD-PIN profile to be caused by Poisson fluctuations.
An independent and bin-free Z$^2_n$-test
	\citep{Buccheri1993AdSpR,Brazier1994MNRAS},
	together with a widely used choice of $n=2$,
	gives $Z^2_n=21.8$ for the HXD data at the period of (\ref{eq:period}),
	with a chance probability of $2.1\times10^{-4}$.
These tests reveal that the HXD-PIN signals are pulsed,
	at the period of equation \ref{eq:period},
	with a high significance ($>$99.9\%),
	even though this level would not be high enough 
	to support the detection of a periodicity without the aid of the XIS data.

Fgure~\ref{fig:pulseprofiles} shows 
	pulse profiles in the 0.8--4.0, 4.0--10, 10--70, and 80--150 keV energy ranges
	observed the XIS (XIS0, XIS1, and XIS3), HXD-PIN, and HXD-GSO data,
	all folded at the period of equation (\ref{eq:period}),
	employing an epoch of 54326.356 (MJD).
Thus, the pulsation is less significant in the 80--150 keV  HXD-GSO data,
	with $\chi^2/\nu=12.2/6=2.0$ (null hypothesis probability 0.06).
The pulse profiles exhibit some energy-dependent changes,
	in agreement with previous reports \citep{Kuiper06,Hartog08}.
The peak-to-bottom pulsed fraction,
	$(F_{\rm max}-F_{\rm min})/(F_{\rm max}+F_{\rm min})$,
	where $F_{\rm max}$ and $F_{\rm min}$ are 
	the maximum and minimum background-subtracted count rates
	across the pulse phase,
	are evaluated to be 
	8.4$\pm$0.3\%,
	13$\pm$1\%,
	and 
	$\sim$41\% 
	in the 0.8--4, 4--10, and 10--70 keV energy ranges,
	respectively.
	
\subsection{Short-Term Variations}
\label{sub:variations}

Further periodogram analyses of the HXD-PIN data
	gave an indication that 
	the pulsations in the 12--40 keV band 
	become more prominent toward the latter part of the observation.
We hence divided all the exposure into three epochs of $\sim$30 ks each,
	and produced three sets of 12--40 keV HXD-PIN light curves
	folded 
	at the period of equation (\ref{eq:period}).
As shown in figure~\ref{fig:compare_pls},
	the pulse profiles from the 1st and  2nd epochs were 
	found to be consistent with each other within statistical errors.
In contrast, 
	as shown in panels (a) and (c) of  figure~\ref{fig:compare_pls},
	those from the 1st and  last epochs appear considerably different;
	the pulses became deeper and more double-peaked in the 3rd epoch,
	while the average intensity decreased
	in agreement with figure~\ref{fig:lc}.1
These effects are unlikely to be caused 
	artificially by cumulative phase errors,
	which is estimated to be $<$9\% of a  cycle
	when folded  at equation (\ref{eq:period}).

To quantitatively examine
	the suggested pulse-profile changes,
	we took phase-by-phase differences between 
	the 1st and 3rd epoch pulse profiles,
	and obtained $\chi^2=20.8$ for $\nu=9$.
The calculation took into account  (in quadrature) 
	the typical systematic background uncertainty  by 0.004 c s$^{-1}$ (1 sigma).
Since the implied null-hypothesis probability is 1.3\%,
	the  two folded profiles are inferred to be different
	at a 98\% confidence level.
This suggests that the 12--40 keV signals changed,
	either in the average intensity or in the pulse profiles (or both).
However, these alternative possibilities cannot be distinguished, 
	since the significance of the profile difference
	reduces to $\chi^2=15.3$ for $\nu=9$ 
	(or a chance probability of $8\%$)
	if the difference profile is fitted by  a constant.
}

For reference, 
	the XIS pulse profiles in 0.4-4 keV did not change
	by more than 7\% among the 3 epochs.
Any spectral change in the XIS or HXD-PIN band was not significant
	among the 3 epochs, either.

\subsection{Spectral Analysis}

\subsubsection{Wide-band spectra}
\label{subsub:XIS+PIN}

To quantify the spectral soft and hard components simultaneously,
	we jointly fitted the data from the XIS, HXD-PIN, and HXD-GSO,
	in the 0.8--10, 12--70 keV, and 50--200 keV ranges, respectively.
Although the signal detection with HXD-GSO 
	is not significant, 
	we include the GSO data points 
	with the systematic error properly added, 
	because we can use them at least as upper-limit points.
The on-source and corresponding background spectra 
	were prepared as stated in section~\ref{sub:bgsubt}.
We also incorporated the correction factor for HXD-GSO spectra, 
	which has been introduced to reproduce 
	the Crab spectra by a single power-law \citep{H.Takahashi08}.

As figure~\ref{fig:rawspec}b shows,
	the XIS and HXD data points happen to define,
	in a rather clear-cut way,
	the soft and hard components, respectively. 
As the simplest attempt, 
	we separately fitted the XIS spectra with an absorbed blackbody model and
	the HXD one by a power-law model,
	to find that the latter is successful with $\Gamma_{\rm h}=0.8^{+0.4}_{-0.2}$
	(the former is not; see below).
As a next attempt,
	we tried to jointly reproduce the whole data simultaneously 
	by an absorbed blackbody model
	plus a power-law model,
	representing the soft and hard components, respectively.
The model normalization between the XIS and HXD-PIN 
	was constrained to be 1.18 \citep{Maeda2008}.
When the hard power-law photon index 
	is fixed, for simplicity, 
	at the above value of $\Gamma_{\rm h} = 0.8$,
	this model gave 
	a blackbody temperature of $kT \sim 0.46$ keV.
However, 
	as shown in figure~\ref{fig:specfit}c,
	 the fit was unsuccessful with $\chi^2/\nu=7326.9/412=17.8$,
	due to significant residuals from the blackbody model 
	at $\sim$1.2 keV and $\sim$5 keV.
The resultant parameters are shown in table~\ref{tab:fit} as Model A.	
When $\Gamma$ is allowed to be vary,
	it became very steep ($\Gamma \sim 3.7$)
	so as to account for the model inaccuracy below $\sim$10 keV
	rather than to reproduce the hard tail.

Trying to improve the fit,
	we added a second steep power-law,
	so that the soft component is now expressed by 
	a sum of a blackbody and a power-law (i.e., BB+PL).
The obtained best-fit solution is described by
	a blackbody of $kT=0.43$ keV,
	a steep power-law of $\Gamma=4.0$,
	and a hard power-law of $\Gamma_{\rm h}=0.1$
	(which is now left free),
	with a three times larger $N_{\rm H}$ value than Model A.
Systematic errors of $\Gamma_{\rm h}$ are estimated to be $\sim$30\% 
	when considering the 2\% uncertainty of the PIN-NXB.
However,
	as shown in figure~\ref{fig:specfit}d (Model B in table~\ref{tab:fit}),
	 this model was still unsuccessful 
	with  $\chi^2/\nu=589.9/409=1.44$,
	mainly due to deviations in the low energy range of the XIS.

By replacing 
	the steep power-law with a second blackbody,
	and 
	thus representing 
	the soft component by two blackbodies (or 2BB),
	the fit was significantly improved to $\chi^2/\nu=459.8/409=1.12$;
	these results are shown in figure~\ref{fig:specfit}e (Model C in  table~\ref{tab:fit}).
The two blackbody temperatures were obtained as 
	$kT_1=0.34$ keV and 
	$kT_2=0.63$ keV,
	with their emission radii (assuming circular regions) as
	13.2 km and 2.3 km, respectively,
	when assuming a distance to the source as 3.6 kpc \citep{Durant06a}.
The absorbing column became
	$N_{\mathrm{H}}=6.4\times 10^{21}$ cm$^{-2}$.

Although the above Model C is approximately successful,
	the fit still leaves significant residuals.
As one of alternative empirical models for the soft component,
	we employed,
	after \citet{Tiengo2005A&A} and \citet{Enoto2010ApJL},	
	 the Comptonized blackbody model 
	described in \citet{Enoto2010PASJ}.
As shown in figure~\ref{fig:specfit}e and
	summarized as Model D in table~\ref{tab:fit},
	this model gave a better fit;
	$\chi^2/\nu=441.5/410=1.08$,
	with 
	$kT = 0.29$ keV,
	a soft power-law photon index $\Gamma_{\rm s}=3.8$,
	and 
	$\Gamma_{\rm h}$=0.4.
The results using this particular model
	was already reported previously \citep{Enoto2010ApJL}.

In the strong magnetic field as $\sim$$10^{14}$ G,
	resonant cyclotron scattering of soft X-ray photons
	becomes more efficient than Thomson scattering.
This effect can be expressed by 
	``Resonant Cyclotron Scattering (RCS)'' model 
	\citep{Thompson02,Lyutikov06,Rea08},
	which calculates a blackbody spectrum
	modified by resonant cyclotron up-scatterings 
	off hot electrons in the neutron star magnetosphere.
This model, combined with the hard power-law,
	has given a still better account of the data 
	over the 0.8--200 keV energy band,
	with $\chi^2/\nu=362.0/409=0.88$.
This fit is shown in figure~\ref{fig:specfit}a and 
	figure~\ref{fig:specfit}f, and in table~\ref{tab:fit} as Model E.
The derived blackbody temperature ($\sim$0.3 keV) is similar to
	that of Model E,
	while the Compton optical depth,
	which was left implicit in Model D, became $\tau \sim 2.0$.
The photon index of the hard power-law became 
	$\Gamma_{\rm h}=0.89_{-0.10}^{+0.11}(\textrm{stat.})^{+0.11}_{-0.10}(\textrm{sys.})$,
	which is close to the value of $\Gamma_{\rm h}\sim0.80$
	obtained using the HXD-PIN data only.

Figure~\ref{fig:nuFnu} represents
	the $\nu F_{\nu}$ spectra of 4U 0142+61,
	fitted with the most successful RCS plus power-law model (Model E) described above.
The X-ray flux in the 10--70 keV energy band is 
	$4.4\times 10^{-11}$ erg cm$^{-2}$ s$^{-1}$,
	while the unabsorbed 1--10 keV flux is 
	$1.8\times 10^{-10}$ erg cm$^{-2}$ s$^{-1}$.
If the hard component extends to 200 keV,
	its flux in the 10-200 keV band
	becomes $1.5\times 10^{-10}$ erg cm$^{-2}$ s$^{-1}$,
	which is comparable to that of the soft component.

\medskip
\subsubsection{Possible local spectral features}
\label{subsub:localfeatures}

At a closer inspection,
	we still find negative residuals at $\sim 20$ keV and $\sim40$ keV
	in the HXD-PIN spectrum (figure~\ref{fig:rawspec}).
Since the NXB spectrum of HXD-PIN shows no spectral features at $\sim 20$ keV,
	and only a very weak Gd-K$_{\alpha}$ emission line at $\sim 42$ keV,
	the suggested features cannot be instrumental.
However, we refrain from further analysis of these features,
	because their statistical significance is not high enough.

Lastly, we examined the hard power-law for a high-energy cutoff,
	because 
	the hard-tail component must steepen at some energies,
	in order for its luminosity not to diverge. 
For this purpose,
	we employed Model D, 
	and multiplied its hard power-law with an exponential cutoff factor 
	of the form $\exp (-E/E_{\rm cut})$, 
	where $E$ is the photon energy and $E_{\rm cut}$ is a cutoff parameter. 
Then, the data gave a lower limit as $E_{\rm cut}  >$ 180 keV at the 90\% confidence limit.
The choice of Model D is because 
	it gives (among the successful 3 models)
	the flattest value of $\Gamma_{\rm h}$,
	and hence the derived lower limit on $E_{\rm cut}$ 
	is considered to be most conservative. 

%

\section{Discussion}


\subsection{Wide-band spectra of magnetars}
Recent X-ray observations of magnetars show that 
	their persistent broad-band spectra
	are commonly represented by 
	an optically thick soft component of $kT \sim 0.3$ keV,
	and 
	a power-law-like hard-tail component 
	which dominates in energies above $\sim$10 keV.
Besides, 
	it was revealed that 
	the two-component spectrum 
	depends significantly on their characteristic age, 
	in such a way that 
	the hard-tail component 
	becomes weaker (relative to the soft component),
	yet harder, towards sources with older characteristic ages
	\citep{Kaspi2010ApJ, Enoto2010ApJL}.
When systematically studying such spectral properties,
	including in particular the evolutionary behavior 
	revelead by \citet{Enoto2010ApJL},
	4U~0142+61 is considered to play a particular important role,
	for the following three reasons.
	
First, among the magnetars in the evolutionary study by \citet{Enoto2010ApJL},
	4U~0142+61 has the oldest characteristic age, 70 kyr, 
	except for 1E~2259+685
	of which the hard-tail component is not well constrained.
As a result,
	4U~0142+61 is currently the oldest magnetar with 
	the well determined hard-tail component,
	and hence it significantly reinforces the age-correlated spectral evolution of magnetars
	 \citep{Enoto2010ApJL}.
Secondly, 
	as a common characteristic of aged magnetars \citep{Enoto2010ApJL},
	the two spectral components of 4U~0142+61 are well separated from each other 
	(see figure~\ref{fig:rawspec}).
This, combined with the relatively high X-ray intensity of 4U~0142+61,
	allows us to better quantify its two spectral components than 
	in other objects of this class.
Finally, the extreme flatness ($\Gamma_{\rm h}\sim 0.8$) 
	of the hard tail of 4U~0142+61,
	which is again typical of aged magnetars,
	is expected to provide very strong constraints on the emission mechanism of 
	the magnetar hard X-rays.
		
\subsection{Soft X-ray emission of 4U~0142+61}


When comparing among the soft X-ray spectral models in \S~\ref{subsub:XIS+PIN},
	the blackbody plus soft power-law model (Model B)
	requires a higher column density of photo-absorption 
	than the other models.
This is likely to be an artifact, 
	necessitated by the too steep low-energy rise of the soft power-law. 
This 	soft power-law component, 
	in addition,
	causes an infrared divergence
	after eliminating the photo-absorption.
Further considering the unacceptable reduced chi-square value,
	this Model B is unlikely to represent the soft component.
Although the two-blackbody model (Model C) 
	can gives a better fit,
	the chi-squared value is still somewhat large.
In addition, 
	the associated hard-tail shape,
	$\Gamma_{\rm h}\sim 1.5$, is considerably larger than
	that obtained with the HXD alone.
	
In contrast to Model B and Model C	
	which employ an ad-hoc combination of simple model components,
	the two more physical models,
	Model D and Model E, were found to be more successful
	with reduced chi-square values of 1.08 and 0.88, respectively
	(table~\ref{tab:fit}).
Both models assume
	a scenario 
	that some seed photons are repeatedly up-scattered by hot electrons in the stellar magnetosphere;
	the former is empirically defined \citep{Enoto2010PASJ},
	while the latter is based on a simplified 1D semi-analytical modeling \citep{Rea08}.
Thus, these successful wide-band fits suggest
	that the known deviations of the soft component 
	from a simple blackbody can be ascribed, at least partially, to  
	up-scattering of soft seed photons off hot electrons located somewhere.	
Below,	
	let us discuss a possible scenario,
	employing the latter modeling (Model E),
	which is more specific.

As shown in fiture~\ref{fig:specfit},
	Model~E has given an acceptable fit to the 0.8--200 keV spectrum,
	with a seed-photon temperature of $kT=0.30$,
	an optical depth of $\tau=2.0$, 
	and an electron thermal velocity of $\beta=0.3$.
These parameters are consistent 
	with $kT=0.30$, $\tau=1.9$, and $\beta=0.33$,
	derived by a combined fitting 
	of the 2004 XMM-Newton observation and 2003-2006 INTEGRAL observations 
	using the same RCS plus a hard tail power-law modeling \citep{Rea08}.
Thus the soft X-ray spectral shape measured in 2007 has little difference from that in 2004.
Therefore,
	the persistent soft X-ray emission of 4U~0142+61
	is relatively stable at least on a time scale of a few years.
In addition,
	the absorbing column density measured with Suzaku,
	$N_{\mathrm{H}}=6.72_{-0.04}^{+0.38}\times 10^{21}$ cm$^{-2}$,
	is in a good agreement with 
	$N_{\mathrm{H}}=(6.4\pm 0.7)\times 10^{21}$ cm$^2$ 
	derived by absorption edges of elements 
	such as O, Fe, Ne, Mg, and Si \citep{Durant06b}.

When using Model E, 
	the overall soft-component shape mainly determines the blackbody temperature $kT$,
	while the data excess above the blackbody toward higher energies,
	to be called ``soft tail", specifies $\tau$ and $\beta$. 
The derived blackbody  temperature, $kT\sim 0.3$ keV, 
	is similar to those derived by Model C and Model D.
(If using a more familiar Comptonized blackbody model, ``compbb",
	we obtain $\tau \sim 2.9$, $kT\sim 0.32$ keV, 
	and an electron temperature $kT_{\rm e}\sim 2.3$ keV.)
Then, 
	let us discuss 
	whether these conditions of Model E are physically realistic in view of general pictures of 
	magnetized neutron stars.
These parameters imply that 
	seed photons with $kT \sim 0.3$ keV,
	presumably from the stellar surface,
	pass through a scattering slab with an optical depth of $\sim 2$,
	filled with electrons which have typical velocities of $0.3c$,
	where $c$ is the speed of light.
Since the magnetic field of 4U~0142+61 
	is estimated as $B \sim 1.3 \times 10^{14}$ G
	from the period $P\sim 8.68$ s and its derivative $\dot{P} \sim2 \times 10^{-12}$ s s$^{-1}$,
	the corresponding Goldreich-Julian density \citep{Goldreich1969ApJ} becomes
	$n_{\rm GJ} =  7\times 10^{13}  (P/1 {\rm \ s})^{-1} (B/10^{15} {\rm \ G}) {\ \rm cm^{-3}}=1\times 10^{12}$
	cm$^{-3}$.
If assuming this density, the Thomson cross-section $\sigma_{\rm T}$,
	and a slab thickness of $l\sim 10^7$ cm,
	the corresponding optical depth becomes
	$\tau \sim n_{\rm GJ} \sigma_{\rm T} l \sim 7 \times 10^{-6}$,
	which is much smaller than the value, $\tau \sim 2$, 
	required by the soft-tail feature in the data.
However,
	consideration of more detailed physics of magnetars,
	as employed in the RCS model,
	allows us to solve this discrepancy by invoking two effects;
	a larger electron density and 
	an enhanced electron-photon cross-section.	
The former is expected under a twisted magnetic configuration,
	where electric currents are induced to give $\sim$3--4 orders of magnitude higher electron density 
	than $n_{\rm GJ}$ \citep{Thompson02}.
The latter, an enhanced cross-section,
	can occur at a cyclotron resonance energy in the scattering slab
	\citep{Lyutikov06}.
These larger density and cross-section 
	are consistent with a basic scheme of the magnetar hypothesis.	


\subsection{Hard X-ray emission of 4U~0142+61}
	
The fit with Model E has given 
	the photon index of the hard component
	as
	$\Gamma_{\rm h}=0.89_{-0.10}^{+0.11}(\textrm{stat.})^{+0.11}_{-0.10}(\textrm{sys.})$.
This is consistent with,
	and of similar quality to,
	the value of $0.93\pm0.06$ measured by INTEGRAL \citep{Hartog08},
	or that based on the XMM-Newton and INTEGRAL data, $1.1\pm0.1$ \citep{Rea08},
	even though the Suzaku exposure, $\sim$100 ks,
	is much shorter than that with INTEGRAL.  
Among the magnetars shown in figure~4 in \citet{Enoto2010ApJL},
	4U~0142+61 thus exhibits one of the hardest power-law components.

The emission mechanism of the hard X-ray component 
	is a topic of extensive discussion.
Although 
	several scenarios have been proposed,
	none of them succeeds to explain all the observational properties.
So far,
	one of the biggest difficulties therein 
	has been 
	how to explain 
	the extremely hard photon indices ($\Gamma_{\rm h} \sim 1$ or less; figure~\ref{fig:nuFnu}).
Now, the difficulty is enhanced by the systematic change in $\Gamma_{\rm h}$
	(from $\sim$1.7 to  $\sim$0.4)
	in correlation with the characteristic age
	\citep{Kaspi2010ApJ,Enoto2010ApJL}.
Indeed,
	this effect 
	cannot be readily explained 
	by the non-thermal radiation model from fast-mode break down \citep{Heyl2005ApJ},
	or 
	the thermal bremsstrahlung model from a transition layer \citep{2005ApJ...634..565T, Beloborodov07},
	or 
	the resonant inverse-Compton up-scattering model \citep{Baring07}.

A possible observational hint to the emission mechanism 
	of the hard component may be provided by the highest energies 
	to which the hard tail extends.
There is a relatively strong COMPTEL upper limit above $\gtrsim$750 keV \citep{Kuiper06}.
In order to make this point clear,
	we show, in figure~\ref{fig:nuFnu_wide},
	a multi-band spectral energy distribution of 4U~0142+61,
	incorporating previously reported multi-wavelength observations.	
Thus,
	if time variations are negligible among these observations,
	the hard X-ray spectrum must turn over at energies between $\sim$200 and $\sim$750 keV.
The Suzaku lower limit on the cutoff, $E_{\rm cut} > 180$ keV, is consistent with 
	the above estimate.
If this hard power-law component originates from 
	non-thermal emission from accelerated particles,
	the cutoff would reflect the maximum particle energy.
However,
	considering the difficulty with the ordinary non-thermal emission 
	processes by energetic particles to explain 
	the values of $\Gamma_{\rm h}$ and its systematic source-to-source differences,
	the cutoff may alternatively be interpreted as 
	an effect of the extremely high magnetic field,
	typically exceeding the critical field $B_{\rm cr}=4.41\times 10^{13}$ G,
	where gamma-rays cannot directly escape 
	due to photon splitting \citep{2006RPPh...69.2631H}.

This photon-splitting view implies a possibility of the hard X-ray emission 
	arising via down-cascade of energetic gamma-rays \citep{Harding1997ApJ,Enoto2010PhD}.
Near the stellar surface,
	gamma-rays above a few hundred keV can be produced
	through, e.g., resonant cyclotron scattering,
	annihilation of electron-positron pairs at the dense stelar atmosphere,
	and sporadic reconnection.
In ultra-strong magnetic fields exceeding $B_{\rm cr}$,
	electron-positron pair cascades are suppressed \citep{Baring2001ApJ},
	while the photon splitting may be dominant \citep{Baring2001ApJ,2006RPPh...69.2631H}.
As a result,
	these gamma-rays from the surface 
	may repeatedly split into lower energy photons,
	since 
	the photon splitting process has no low-energy threshold
	unlike the one-photon pair production process.
In fact, \citet{Harding1997ApJ} 
	reported that 
	the photon splitting becomes important at $B\gtrsim 0.3B_{\rm cr}$,
	since the attenuation length due to splitting 
	become comparable to or less than those for pair production	.
This process can also explain the differences in $\Gamma_{\rm h}$ among magnetars,
	in such a way that
	higher fields of younger objects will allow the photon-splitting cascade
	to proceed down to lower energies, 
	and hence to make the continuum softer. 
	
\subsection{Energetics}
Assuming an isotropic emission at an distance of 3.6 kpc \citep{Durant06a},
	the soft (RCS) and hard components are implied to have 
	unabsorbed luminosities of 
	$2.8\times 10^{35}$ erg s$^{-1}$ and 
	$6.8\times 10^{34}$ erg s$^{-1}$, 
	in the 1--10 keV and 10--70 keV band, respectively.
If calculated in the 2--10 keV band,
	our soft-component luminosity becomes $1.1 \times 10^{35}$ erg s$^{-1}$.
Since this agrees with the previous measurements \citep{Durant06a},
	the soft component is inferred to be stable,
	making some theoretical prediction \citep{Thompson96}
	consistent with the observation.
Adding up the soft and hard components,
	the absorption-corrected 1-200 keV luminosity becomes 
	$5.2 \times 10^{35}$ erg s$^{-1}$,
	which can be
	decomposed into $\sim 54$\% and $\sim 46$\%
	carried by the soft and hard components, respectively.
Since the spin-down luminosity of the source
	is only $1.2 \times 10^{32}$ erg s$^{-1}$,
	neither components can be powered by the rotational energy.

\subsection{Pulse profiles}
The soft and hard X-ray pulse profiles, 
	obtained with Suzaku (figure~\ref{fig:pulseprofiles}),
	are generally consistent with those 
	measured previously with INTEGRAL and RXTE \citep{Kuiper06, Hartog08}.
In energies below a few keV, 
	the profile has two peaks,
	and one of them (phase at $\sim$0.5 in our figure~\ref{fig:pulseprofiles}
	and ~0.1 in figure~5 of \cite{Kuiper06})
	becomes weaker towards ~10 keV,
	but it partially recovers towards a few tens keV
	at a somewhat smaller pulse phase.

As seen above, 
	the pulse profiles of 4U~0142+61 depend in a complex way on the energy.
However,
	the main peaks of the two spectral components 
	are at the same phase $\sim$1 in figure~\ref{fig:pulseprofiles},
	and their emission regions seem to be located at the same rotation phase.
Regarding the soft component as an optically-thick thermal emission
	from the polar-cap regions of the neutron star,
	we further speculate
	that the spectral hard-tail component,
	which are generally pulsed strongly (e.g., \cite{Kuiper06}),
	is also emitted from the polar-cap regions.
This inference gives an additional support 
	to the photon-splitting mechanism,
	because the input gamma-rays will be produced,
	e.g., via electron-positron annihilation,
	mainly at the polar-cap regions.

	
In subsection 3.4 and figure~\ref{fig:compare_pls}, 
	we presented evidence for short-term (in ~30 ks) variations
	in the pulse profile (or overall signal intensity).
During the last 30 ks of our exposure,
	when the hard X-ray intensity slightly decreased (figure 1),
	the HXD-PIN pulse profile possibly became 
	more strongly pulsed and more double peaked.
These effects, if real, may reflect some sporadic processes
	involved in the hard-tail production mechanism.
For example, if the persistent emission of a magnetar is formed 
	by an assembly of numerous small short bursts \citep{Nakagawa2007PASJ},
	the persistent intensity should fluctuate 
	due to statistical fluctuations of the number of small bursts.
We expect the soft spectral component to be more stable,
	because of the heat capacity of the neutron star.
More quantitative evaluation of these issues will be presented elsewhere.

\section{Conclusion}
We observed the prototypical anomalous X-ray pulsar 4U 0142+61 
	with Suzaku, 
	and quantified its wide-band spectra, 
	spanning from 0.4 to $\sim$70 keV or higher (subsection 3.2).
This is a first detailed broad-band study of this source performed 
	using a single satellite within a short exposure ($\sim$100 ks), 
	without cross-calibration uncertainties among multiple satellites. 
We reconfirmed the hard-tail component above 10 keV,
	with high significance of $\gtrsim 3\sigma$.
Detailed analyses of the soft and hard X-ray spectra, 
	simultaneously fitted by several spectral models (subsection 3.5.1),
	shows that
	the soft component is represented by 
	Comptonized blackbody model or Resonant Cyclotron Scattering model
	with a seed-photon temperature of $kT \sim 0.3$ keV.
The hard-tail component can be fitted by a 
	power-law with a photon index of 
	$\Gamma_{\rm h} \sim 0.9$,
	together with a constraint on its cutoff as $E_{\rm cut} > 180$ keV.
The folded pulse profiles in the 12--40 keV range 
	exhibited evidence for shape or intensity variations 
	on a time scale of 30 ks (subsection 3.4),
	suggesting some sporadic process in the hard X-ray production process.
We propose the photon splitting process
	as a promising production mechanism for the hard X-ray component.
	
 \bigskip
We are grateful to all the Suzaku team members, 
	including in particular M. Morii for his elucidating discussion.


\clearpage
\begin{table}
\begin{center}
\caption{Spectral parameters of  4U~0142+61 observed in 2007 August\footnotemark[$*$].}
\label{tab:fit}
\begin{tabular}{lcccccc}
\hline \hline
& Model A & Model B & Model C & Model D & Model E \\
& BB+hPL & BB+sPL+hPL  & 2BB+hPL & CBB+hPL & RCS+hPL \\
\hline
Absorbed soft flux \footnotemark[$\dagger$] &
$66.0$ &
$64.7\pm 0.2$ &
$64.4\pm0.6$ &
$64.4\pm0.2$ &
$64.4^{+7.0}_{-6.4}$  \\
Absorbed hard flux \footnotemark[$\ddagger$] &
$83.2$ &
$34.6^{+3.7}_{-4.0}$&
$21.7^{+3.0}_{-3.8}$&
$37.2^{+3.5}_{-3.7}$&
$32.8^{+3.2}_{-2.9}$ 
\\ 
\hline
 $N_{\rm H}$ ($10^{22}$ cm$^2$) &
$0.40$ &
 $1.12\pm 0.02$ &
 $0.64\pm 0.01$ &
 $0.63\pm0.01$ &
 $0.672^{+0.038}_{-0.004}$ \\
  $kT_1$ (keV) & 
--  &
-- &
$0.337^{+0.006}_{-0.007}$ &
-- & 
--
 \\
 $R_1$ (km)  \footnotemark[$\S$]  &
--  &
-- &
$13.2^{+0.6}_{-0.5}$ &
-- & 
-- 
\\
$kT_2$ (keV) & 
-- &
 -- &
 $0.63\pm 0.02$  & 
 --  &
 --
 \\
  $R_2$ (km)  \footnotemark[$\S$]  &
--  &
  -- &
$2.3^{+0.3}_{-0.2}$ &
  -- &
  -- 
 \\
 $kT$ &
$0.46$  &
$0.428\pm 0.004$ &
  -- &
$0.289^{+0.003}_{-0.002}$ & 
$0.30^{+0.03}_{-0.12}$
  \\
   $\Gamma_{\textrm{soft}}$   &
--   &
$3.95^{+0.04}_{-0.03}$ &
 -- &
 -- &
 --
 \\
  $\alpha$ (=$\Gamma_{\textrm{s}}-1$)   \footnotemark[$\|$] &
-- &      
 -- &
 -- & 
 $3.78^{+0.04}_{-0.03}$ &
 --
 \\
  RCS $\tau$ &
-- & 
-- &
 -- &
 -- & 
$1.98^{+0.07}_{-0.06}$
 \\
  RCS $\beta$ &
--  & 
 -- &
 -- &
 -- & 
$0.32^{+0.01}_{-0.02}$
\\
 $\Gamma_{\rm h}$  &
0.80 (fix) & 
$0.11^{+0.13}_{-0.11}$ &
$1.54\pm0.13$ &
$0.40^{+0.11}_{-0.09}$ &
$0.89^{+0.11}_{-0.10}$
 \\
\hline
$\chi^2$/d.o.f &
7326.9/412& 
589.9/409 & 
459.8/409 &
441.5/410 &
362.0/409 
\\
$\chi^2_{\nu}$ &
17.8& 
1.44 & 
1.12 &
1.08 &
0.88 
\\
Probability &
$<10^{-5}$& 
$1\times 10^{-8}$ &
0.042 &
0.14 &
0.95
\\
\hline
\multicolumn{4}{@{}l@{}}{\hbox to 0pt{\parbox{140mm}{
	\footnotesize
	\footnotemark[$*$] 
BB, PL, CBB and RCS represent
	blackbody, powe-law, (magnetic) comptonized blackbody, 
	and 
	resonant cyclotron scattering, respectively.
All the quoted errors are only statistical at the 90\% confidence level.
Lower-case characters ``h" and ``s" in model names 
	represent 
	``hard" and ``soft",
	respectively.
       \par\noindent
       \footnotemark[$\dagger$] 
       The absorbed X-ray fluxes (10$^{-12}$ ergs s$^{-1}$ cm$^{-2}$ ) in the 2--10 keV.
       \par\noindent
       \footnotemark[$\ddagger$] 
       The X-ray flux ($10^{-12}$ ergs s$^{-1}$ cm$^{-2}$ ) in the 15--60 keV. 
       \par\noindent
        \footnotemark[$\S$] Distance to 4U~0142+61 is assumed at 3.6 kpc.
       \par\noindent
       \footnotemark[$\|$]  The parameter $\alpha$ is related with the soft tail power-law 
       	as $\alpha = \Gamma_{\rm s}-1$,
	and describes Comptonization process.
	Details are given in an appendix of \citet{Enoto2010PASJ}.
     }\hss}}
  \end{tabular}
 \end{center}
 \end{table}

\clearpage

\begin{figure}[htb]
\begin{center}
\FigureFile(100mm,){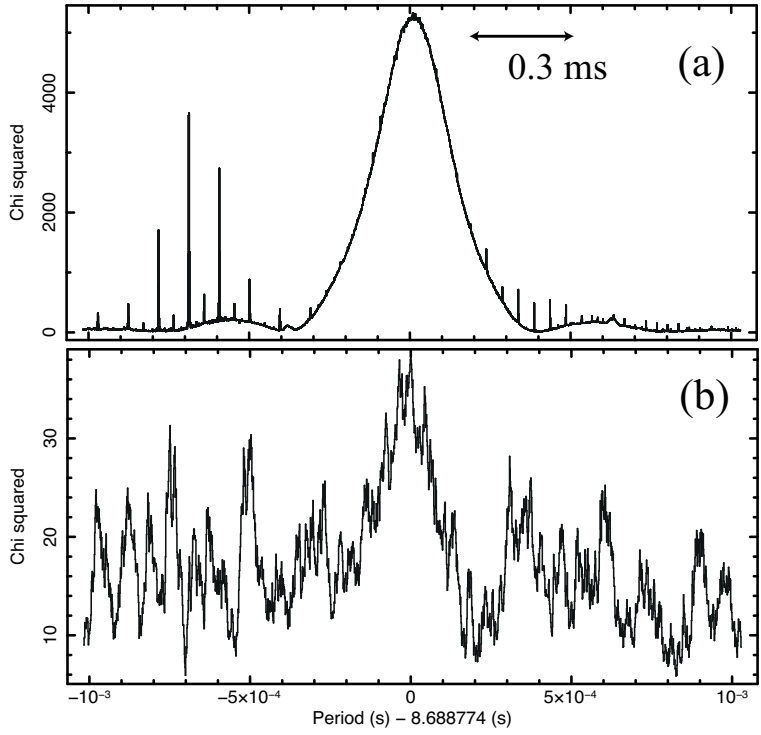}
\end{center}
\caption{
Periodograms from XIS0+XIS1+XIS3  (panel a: 0.8--10 keV)
	and HXD-PIN (panel b: 12--40 keV),
	calculated using 0.54 s binned light curves and 
	16 bins for one pulse period.
The backgrounds are inclusive.
}
\label{fig:periodogram}
\end{figure}

\begin{figure}[htb]
\begin{center}
\FigureFile(80mm,80mm){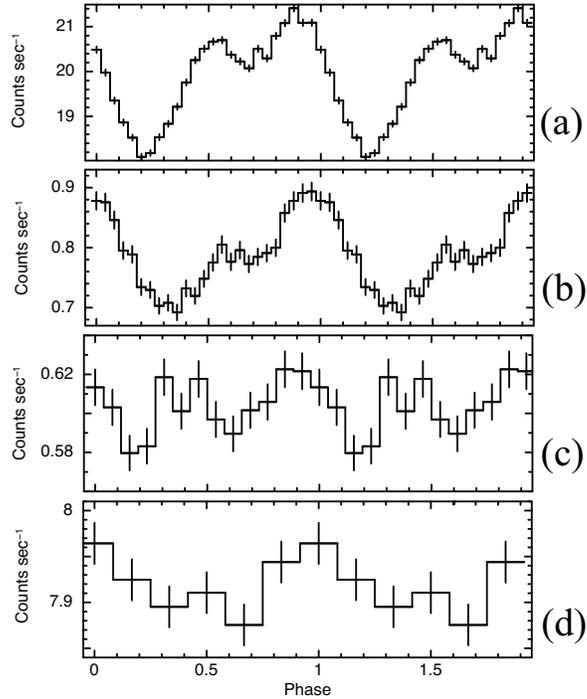}
\end{center}
\caption{
Background-inclusive XIS (XIS0, XIS1, and XIS3 summed), 
	HXD-PIN, and HXD-GSO pulse profiles,
	obtained by folding the total data at the pulse period
	of equation~(\ref{eq:period}),
	in (a) 0.8--4.0, (b) 4.0--10, (c) 10--70, and (d) 80--150 keV ranges.
The HXD data is corrected for the dead time.
Corresponding background rates are estimated to be
	 $(1.34\pm 0.04)\times 10^{-1}$ counts sec$^{-1}$,
	$(1.94 \pm 0.14)\times 10^{-2}$ counts sec$^{-1}$,
	$(5.49\pm 0.11)\times 10^{-1}$ counts sec$^{-1}$,
	and 
	$5.00 \pm 0.21$ counts sec$^{-1}$,
	for panels (a), (b), (c), and (d), respectively.
These background rates include only statistical errors for the XIS
	and
	both statistical and systematic errors for the HXD.
We assumed 2\% and 0.6\% systematic errors for the PIN-NXB and the GSO-NXB, respectively. 
}
\label{fig:pulseprofiles}
\end{figure}

\begin{figure}[htb]
\begin{center}
\FigureFile(80mm,80mm){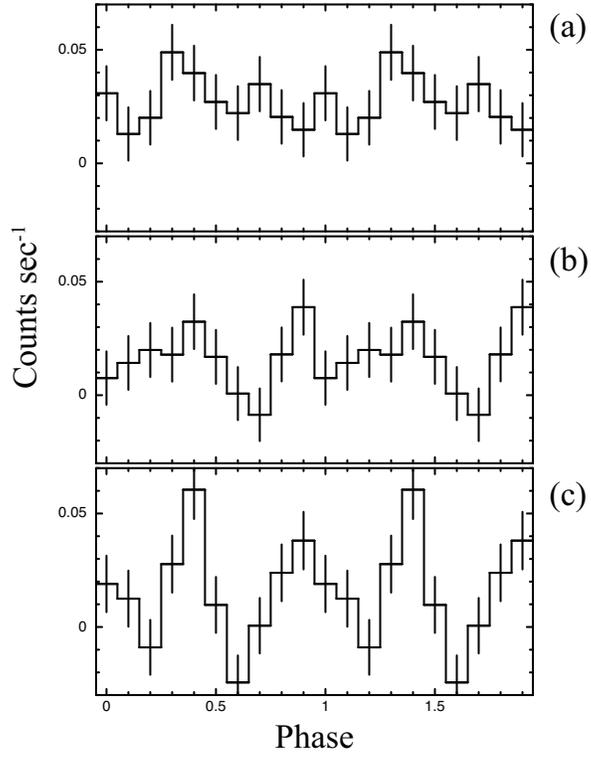}
\end{center}
\caption{
NXB- and CXB- subtracted and deadtime-corrected 
	12--40 keV HXD-PIN pulse profiles,
	during (a) 1st,  (b) 2nd, and (c) 3rd 30 ks intervals, respectively.
Error bars are statistical (1$\sigma$) only,
	whereas the sytematic errors of the background subtraction is 
	$\sim$0.004 c s$^{-1}$.
}
\label{fig:compare_pls}
\end{figure}

\begin{figure}[hbt]
\begin{center}
\FigureFile(100mm,100mm){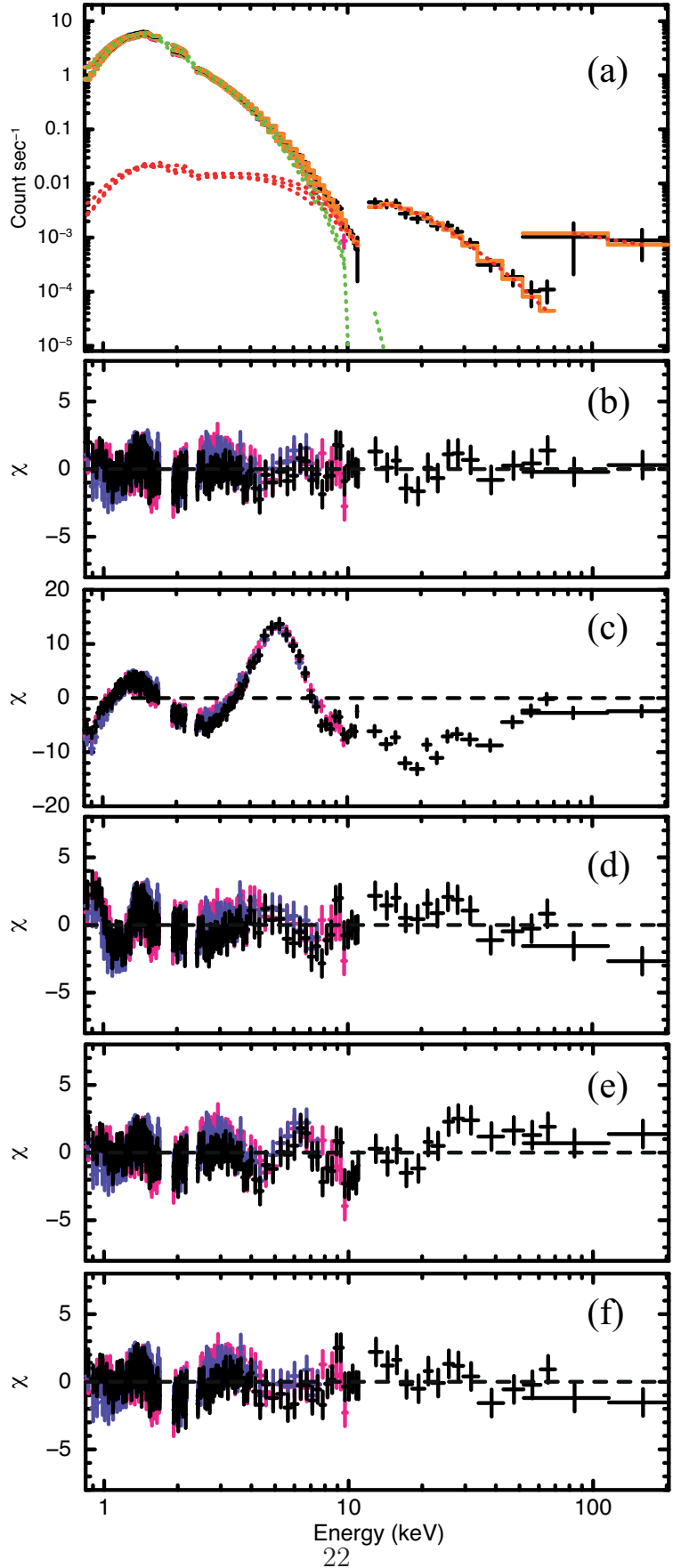}
\end{center}
\caption{
(a)
The same Suzaku spectra of 4U 0142+61  
	as shown in figure~\ref{fig:rawspec}, 
	fitted by RCS (green)+PL (magenta) model, namely Model E.
(b) Residuals from the fit in panel (a).
(c)(d)(e)(f) The same as panel (b), 
	but the soft component is represented by
	BB (Model A), 
	BB+PL (Model B), 
	2BB (Model C),
	and CBB (Model D) models, respectively.
}
\label{fig:specfit}
\end{figure}

\begin{figure}[hbt]
\begin{center}
\FigureFile(130mm,130mm){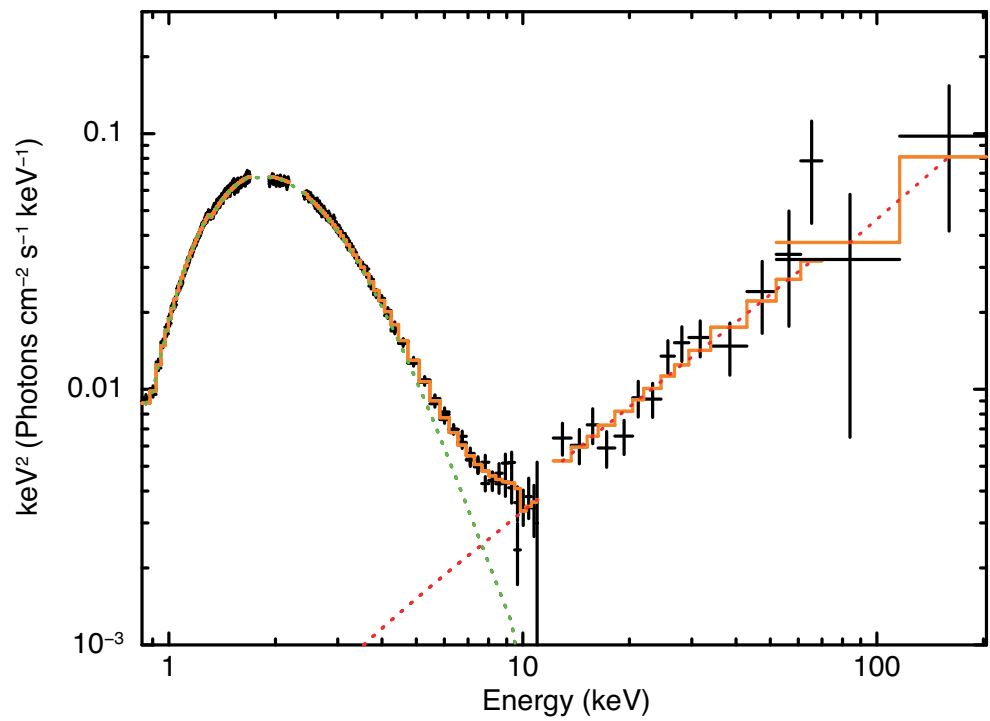}
\end{center}
\caption{
The same results as in figure~\ref{fig:specfit}a
	but shown in the decomvolved $\nu F_{\nu}$ form.
}	
\label{fig:nuFnu}
\end{figure}

\begin{figure}[hbt]
\begin{center}
\FigureFile(160mm,150mm){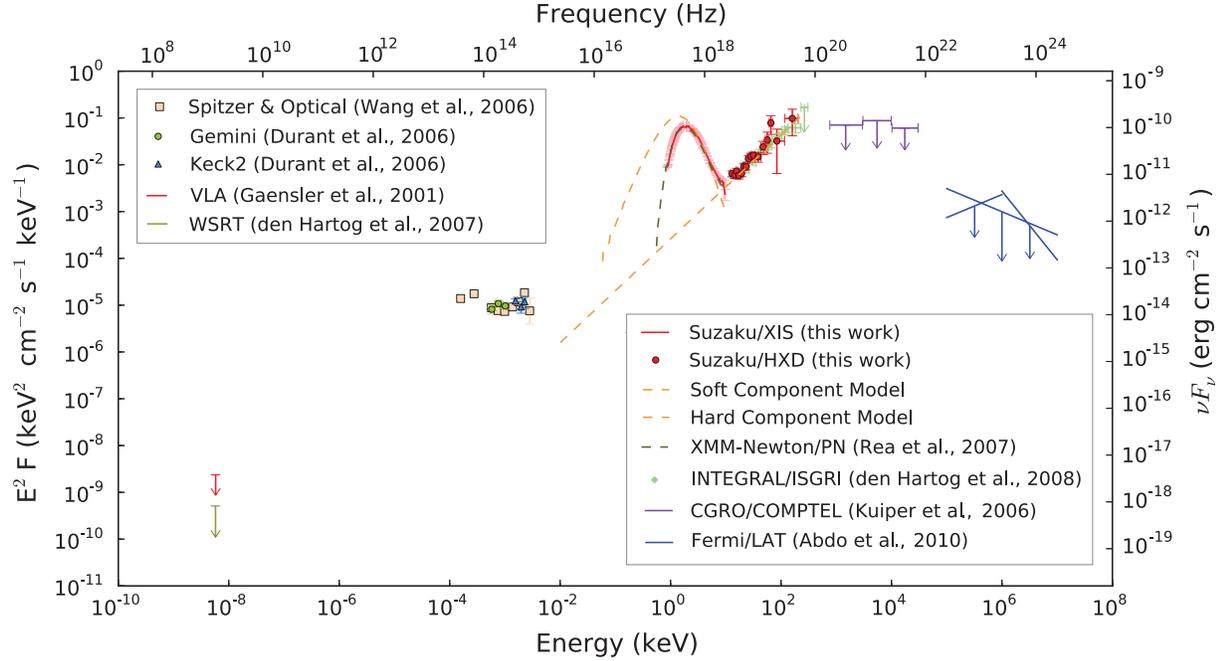}
\end{center}
\caption{
A multi-band spectral energy distribution 	of 4U~0142+61.
The Suzaku data are shown 
	in the 0.8--10 keV and 12-200 keV energy ranges  
	from the XIS and the HXD, respectively.
The soft and hard components of Model~E are represented by dashed lines 
	after corrections for the interstellar absorption,
	and are extrapolated to lower energies.
The previous X-ray results are also shown;
	the 0.1--200 keV band
	from the XMM-Newton/PN \citep{2007MNRAS.381..293R,2010arXiv1011.0091T}
	and 
	the INTEGRAL/ISGRI data sets \citep{Hartog08}.
In the gamma-ray energy band,
	the $2\sigma$ CGRO/COMPTEL upper limits \citep{denHartog2006,Kuiper06} 
	and 
	95\% Fermi/LAT upper limits \citep{2010arXiv1011.0091T}
	are ploted.
In the infrared and optical ranges,
	Spitzer observations at 4.5 $\mu$m and 8.0 $\mu$m \citep{2006Natur.440..772W}
	and 
	some data sets from Gemini and Keck II \citep{Durant06c,2004A&A...416.1037H}
	are shown,
	where the optical data are de-extincted assuming a reddening value of $A_V=3.5$.
In the radio band,
	a 2$\sigma$ upper limit from 1.38 GHz WSRT continuum \citep{2007Ap&SS.308..647D},
	and the 1.4 GHz VLA upper limit \citep{2001ApJ...559..963G} are shown.
}	
\label{fig:nuFnu_wide}
\end{figure}

\end{document}